\documentclass[aps,pra,twocolumn,showpacs,floatfix,superscriptaddress]{revtex4}

\pdfoutput=1

\usepackage{amsmath}
\usepackage{amssymb}
\usepackage{graphicx}
\usepackage{epstopdf}

\newcommand{\bra}[1]{\left<#1\right|}
\newcommand{\ket}[1]{\left|#1\right>}

\begin{document}

\title{Gauge fields emerging from time reversal symmetry breaking   for spin-5/2 fermions in a honeycomb lattice}
\author{G. Szirmai}
\affiliation{ICFO-Institut de Ci\`encies Fot\`oniques, Mediterranean Technology Park, 08860 Castelldefels (Barcelona), Spain}
\affiliation{Research Institute for Solid State Physics and Optics, H-1525 Budapest P.O. Box 49, Hungary}
\author{E. Szirmai}
\affiliation{ICFO-Institut de Ci\`encies Fot\`oniques, Mediterranean Technology Park, 08860 Castelldefels (Barcelona), Spain}
\affiliation{Research Institute for Solid State Physics and Optics, H-1525 Budapest P.O. Box 49, Hungary}
\author{A. Zamora}
\affiliation{ICFO-Institut de Ci\`encies Fot\`oniques, Mediterranean Technology Park, 08860 Castelldefels (Barcelona), Spain}
\author{M. Lewenstein}
\affiliation{ICFO-Institut de Ci\`encies Fot\`oniques, Mediterranean Technology Park, 08860 Castelldefels (Barcelona), Spain}
\affiliation{ICREA-Instituci\'o Catalana de Recerca i Estudis Avan\c cats, Lluis Companys 23, 08010 Barcelona, Spain}

\begin{abstract}
We propose an experimentally feasible setup with ultracold
alkaline earth atoms to simulate the dynamics of U(1) lattice
gauge theories in 2+1 dimensions with a Chern-Simons term. To this
end we consider the ground state properties of spin-5/2 alkaline
earth fermions in a honeycomb lattice. We use the Gutzwiller
projected variational approach in the strongly repulsive regime in
the case of filling 1/6. The ground state of the system is a
chiral spin liquid state with $2\pi/3$ flux per plaquette, which
spontaneously violates time reversal invariance. We demonstrate
that due to the breaking of time reversal symmetry the system
exhibits quantum Hall effect and chiral edge states. We relate the
experimentally accessible spin fluctuations to the emerging gauge
field dynamics. We discuss also properties of the lowest energy
competing orders.
\end{abstract}

\date{\today}
\maketitle

One of the main motivations of studying
ultracold atoms in optical lattices is the high extent of
experimental control. Such systems are very flexible and therefore
are good candidates for simulating other quantum systems, where
experimental control is more cumbersome. There is a vast number of
proposals where ultracold quantum gases can serve as simulators of
condensed matter, or even high energy physics systems (see for
instance Ref. \cite{lewenstein07a}). An important example of such
proposals concerns the recent experimental realization of trapping
and cooling of ultracold alkaline earth atoms
\cite{fukuhara07a,desalvo10a,taie10a}, which could serve for
quantum simulators of high symmetry magnetism \cite{gorshkov10a}.
Despite of these spectacular developments, one of the most
important goals of quantum simulators remains  still to be
realized, namely the simulation of quantum gauge theories, which
appear first of all in high energy physics, but arise naturally
also in many areas of condensed matter physics, such as physics of
frustrated systems, or  of high temperature superconductors
\cite{lee06a}. The main difficulty here is to map the many modes
of the gauge field to those of an atomic ensemble. Very recent
proposals use mixtures of fermionic and bosonic atoms, so that the
bosons are the mediators of the gauge field
\cite{cirac10a,kapit11a}.

Here we propose another, somewhat simpler, scheme with only a
single species of ultracold atoms to simulate a 2+1 dimensional
U(1) lattice gauge theory with a Chern-Simons term. Our proposal
is based on the observation that low energy excitations of certain
Mott insulators can be described by lattice gauge theories
\cite{lee06a,wenbook}. The Mott insulator we consider here is
formed by spin-5/2 alkaline earth atoms, such as
$^{173}\mathrm{Yb}$, which, as was shown by Hermele {\it et al.}
\cite{hermele09a}, can exhibit time reversal symmetry breaking,
and have a so called chiral spin liquid (CSL) ground state in a square
lattice. CSL states lack any kind of long range
order, but due to the violation of time reversal invariance,
they are stable also at low temperatures. The fluctuations above
the CSL state are described by a U(1) gauge theory with a
Chern-Simons term arising from the chiral (time reversal symmetry
breaking) nature of the ground state \cite{wen89a}. Here we treat
the case of a honeycomb lattice; we show that the lowest energy spin liquid ansatz with one
particle per site that respects the underlying lattice symmetries is a CSL. We
describe also two other  lowest energy spin liquid states, that
can be stabilized in certain situations.  We show also how the
dynamics of the emerging gauge fields is measurable by spin
correlation functions.
Obviously, further motivation to study spin
liquid phases in a honeycomb lattice with ultracold atoms is due
to a newly rising interest in related challenging problems, like
superconductivity in graphene
\cite{meng10a,black-schaffer07a,pathak10a,li11a}, or other forms
of  time reversal symmetry breaking that appear for honeycomb and
pyrochlore lattices \cite{machida09a,koch10a}.

The main advantage of using alkaline earth atoms is that the
nuclear spin $I$ decouples from the angular momentum of the
electrons $J$, which is zero in the ground state, and hence
collision processes become spin independent. Therefore spin-$I$
isotopes of alkaline earth atoms can be described to a very good
accuracy by SU($N$) symmetric model Hamiltonians, where
$N=2\,I+1$, and we take $I=5/2$, which is realized e.g. by
$^{173}\mathrm{Yb}$. In an optical lattice the SU($N$) symmetric
Hubbard Hamiltonian takes the form
\begin{equation}
\label{eq:hamhub}
H=-t\sum_{\left\langle i,j\right\rangle,\alpha}\Big(c_{i\alpha}^\dagger c_{j\alpha}+\mathrm{H.c.}\Big)+\frac{U}{2}\sum_{i,\alpha,\beta}c_{i\alpha}^\dagger c_{i\beta}^\dagger c_{i\beta} c_{i\alpha},
\end{equation}
where $c_{i\alpha}$ annihilates an atom at site $i$ with spin
$\alpha\in\lbrace-\frac{5}{2}\ldots \frac{5}{2}\rbrace$, $t$
stands for the tunneling amplitude, and $U$ for the strength of
the on-site interaction.

In the strongly repulsive regime, $U\gg t$, the motional (charge)
degree of freedom of the fermions gets frozen at low temperatures
leading to a Mott insulator state, and the system can be described
by an effective SU(6) spin Hamiltonian \cite{gorshkov10a,wenbook}.
The ground state of such a Hamiltonian can be a N\'eel
antiferromagnetic state with long range order, or a spin liquid
state without any kind of long range order \cite{wen02a} that can
be described using the (resonating) valence bond concept
\cite{anderson73a,fazekas74a,marston89a,lee06a}. Hermele {\it et
al.} have shown that the N\'eel order is ruled out by energy
constraints for SU($N$) Hamiltonians for $N\ge3$
\cite{hermele09a}.

Deeply in the Mott insulator region
  ($U\gg t$) and for 1/6 filling, there is exactly one particle per site. From experimental point
 of view this is an appealing setup, in which undesired 3-body losses can be neglected.
 In this case particle tunneling is forbidden due to the very high energy cost of the multiply occupied sites.
 Only virtual hoping is allowed, and the Hamiltonian \eqref{eq:hamhub} can be approximated by
\begin{equation}
\label{eq:hameff}
H_{\text{eff}}=g\sum_{\left\langle i,j\right\rangle,\alpha,\beta}c_{i\alpha}^\dagger c_{j\alpha}c_{j\beta}^\dagger c_{i\beta},
\end{equation}
with $g=-4 t^2/U$. This effective Hamiltonian acts in the
restricted Hilbert space of 1 atom per site; this condition is
enforced by the local constraints: $\sum_\alpha
c^\dagger_{i\alpha}c_{i\alpha}=1$. The Hamiltonian
\eqref{eq:hameff} already exhibits local U(1) gauge invariance,
since it is invariant under the transformation
$c_{i\alpha}\rightarrow c_{i\alpha}\ e^{i\theta_i}$, in
accordance with the local constraints.

\begin{figure}[tb]
\begin{center}
  \includegraphics{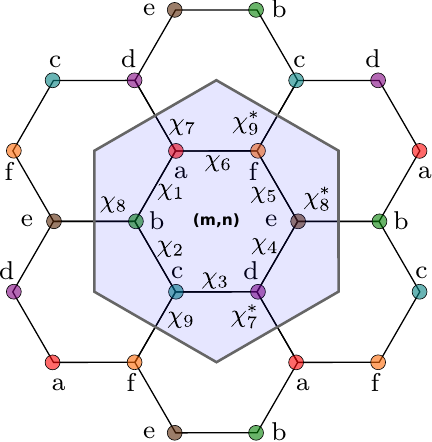}
  \caption{(Color online) The unit cell for the mean-field approximation.}
  \label{fig:unitcell}
\end{center}
\end{figure}
To study the ground state properties and the low energy
excitations of the system we decouple the quartic terms via a
mean-field treatment by introducing the average of the
pair-correlation defined as
\begin{equation}
\label{eq:meanvalue}
\chi_{ij}\equiv\sum_\alpha\langle c_{i\alpha}^\dagger c_{j\alpha}\rangle=\chi_{ji}^*.
\end{equation}
The effective Hamiltonian takes then the mean-field form
\begin{equation}
\label{eq:mfham}
H_{\text{mf}}=g\sum_{\left\langle i,j\right\rangle}\left[\sum_\alpha \left(\chi_{ij} c_{j\alpha}^\dagger c_{i\alpha}+
\chi_{ji} c_{i\alpha}^\dagger c_{j\alpha}\right)-|\chi_{ij}|^2\right].
\end{equation}
Though the mean-field Hamiltonian Eq. \eqref{eq:mfham} is already
quadratic, one still needs to make further assumptions about the
solution in order to obtain a tractable  set of equations. We
choose a hexagonal unit cell containing 6 lattice sites, as
 depicted in Fig. \ref{fig:unitcell}; such cell respects the original
lattice symmetries, and contains as many sites as needed to form a
SU(6) singlet. As a consequence 9 independent mean-field
amplitudes arise: $\chi_1\ldots\chi_{9}$, and the description
through the Hamiltonian \eqref{eq:mfham} inherently describes 6
bands inside the Brillouin zone. With the introduction of 6
Lagrange multipliers (one for each site of the unit cell) the
local constraints can be enforced, and then the Mott insulator
state with 1 particle per site is simply achieved by filling the lowest band.

\begin{figure}[b]
\begin{center}
  \includegraphics{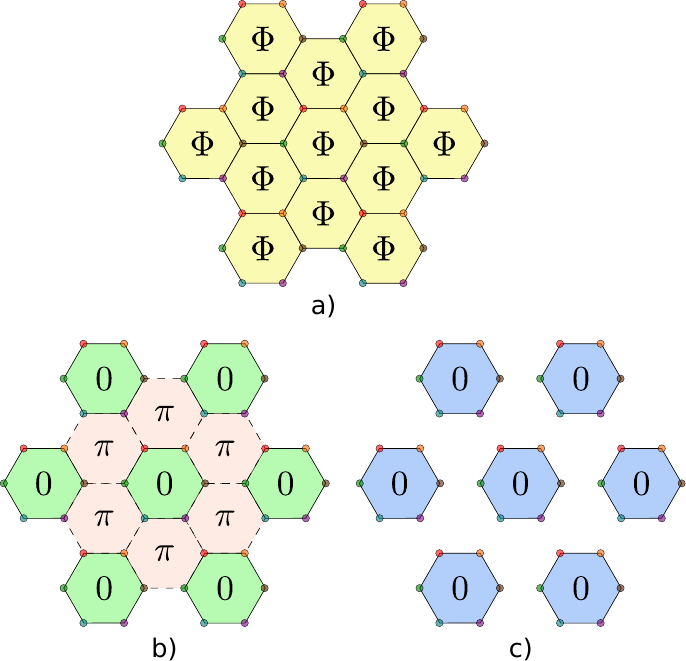}
  \caption{(Color online) Lowest energy mean-field solutions. Subfigure a)
  illustrates the chiral spin-liquid configuration with all bonds having the same
  magnitude. Subfigure b) depicts the quasi plaquette phase with real Wilson loops.
  The dashed line represents a smaller bond value than the solid line.
  Subfigure c) shows the plaquette phase configuration with only one nonzero Wilson loop per three cells.
  Links with a zero mean-field value are removed from the figure.}
  \label{fig:mfsols}
\end{center}
\end{figure}

By diagonalizing the Hamiltonian \eqref{eq:mfham} one obtains the
ground state $\ket{\Psi^{\chi_{ij}}}$; using then the
self-consistency condition \eqref{eq:meanvalue}, one arrives at a
set of equations for the mean-field amplitudes and for the local
Lagrange multipliers. This set of equations is highly nonlinear
and has several solutions. Because of the local U(1) gauge
structure, solutions related to each other by the
transformation
\begin{equation}
\chi_{ij}\rightarrow\chi_{ij}\,e^{i(\theta_i-\theta_j)}
\end{equation}
are equivalent, and have the same energy and the same physical
spin wave function, obtained by the Gutzwiller projection, i.e.,
by restricting the solution to the space with one particle per
site:
\begin{equation}
\Psi(\alpha_1,\alpha_2\ldots\alpha_m)=\bra{0}\prod_i c_{i\alpha_i}\ket{\Psi^{\chi_{ij}}}.
\end{equation}
The Wilson loop $\Pi_i=\prod_{\mathcal{P}_i} \chi_{ij}$ calculated for each plaquette $\mathcal{P}_i$ is invariant under
such transformations, and therefore is the same for gauge equivalent solutions. Note that due to our construction there are
 3 nonequivalent plaquettes that can be considered as a 3 sublattice ansatz on the dual lattice: the triangular lattice
 formed by the plaquettes. Table \ref{tab:mfsol} shows the three lowest energy mean-field solutions.
The ground state solutions (first two lines of Table
\ref{tab:mfsol}) correspond to
$\chi_1=\chi_2\ldots=\chi_6=r\,e^{\pm i\varphi}$ and
$\chi_7=\chi_8=\chi_{9}=r$, with $r\approx0.82651$ and
$\varphi=2\pi/9$. The corresponding Wilson loops are
$\Pi_i=r^6\,e^{\pm i\Phi}$, with $\Phi=2\pi/3$ flux. Therefore the
ground state breaks time reversal invariance, and is a CSL state.
Due to the violation of time reversal symmetry there is a double
degeneracy in addition to the gauge structure. A higher energy
staggered like phase is the next-lowest lying quasi-plaquette
state that has a triple degeneracy with an energy of $E=-6.062$,
as it is shown in the following 3 lines of Table \ref{tab:mfsol}.
This phase is the honeycomb analog of the (staggered) $\pi$-flux
state of spin-1/2 fermions on a rectangular lattice, which is the
lowest energy spin-liquid state for the non-doped Mott insulator.
In our case, due to the frustrated nature of the dual lattice, the
staggered flux phase is not energetically favorable. In this state
the fluxes do not alternate sign, but they change between 0 and
$\pi$ in a way that every zero flux plaquette is surrounded by
plaquettes with $\pi$ fluxes. The last three lines of Table
\ref{tab:mfsol} show a valence bond crystal type ordering,
where there is no flux threading through the disconnected
plaquettes. This state is similar to the ground state of e.g. the
spin-3/2 fermions in the square lattice at quarter filling
\cite{szirmai11a}. This valence bond crystal also has a triple
degeneracy. Figure \ref{fig:mfsols} illustrates these three lowest
energy states.
\begin{table}[tb]
\begin{tabular}{c|c|c|c}
$E$&$\Pi_1$&$\Pi_2$&$\Pi_3$\\
\hline\hline
$-6.148$&$-0.159-0.276i$&$-0.159-0.276i$&$-0.159-0.276i$\\
$-6.148$&$-0.159+0.276i$&$-0.159+0.276i$&$-0.159+0.276i$\\
\hline
$-6.062$&$0.460$&$-0.223$&$-0.223$\\
$-6.062$&$-0.223$&$0.460$&$-0.223$\\
$-6.062$&$-0.223$&$-0.223$&$0.460$\\
\hline
$-6$&$1$&$0$&$0$\\
$-6$&$0$&$1$&$0$\\
$-6$&$0$&$0$&$1$\\
\end{tabular}
\caption{Mean-field solutions. The first column represents the
energy of the mean-field solution, the other three columns give
the Wilson loops of the 3 different plaquettes.} \label{tab:mfsol}
\end{table}

\begin{figure}[t]
\begin{center}
  \includegraphics{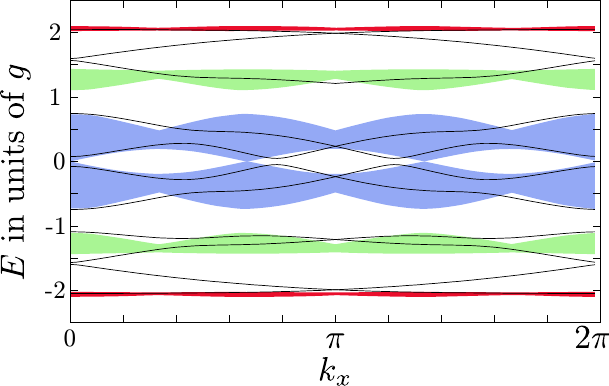}
  \caption{(Color online) The energy spectrum of the system for open boundary conditions in one direction.}
  \label{fig:spectrum}
\end{center}
\end{figure}
In the CSL state the mean-field generated fluxes penetrating the
plaquettes are analogous to those created by a magnetic field. As
a consequence, quantum Hall effect can be observed. It is entirely
generated in this case spontaneously by the mean-fields. It leads
to appearance of chiral edge states, i.e. current carrying states
localized at the boundaries in such a way that the opposite
directions are separated to opposite edges. Figure
\ref{fig:spectrum} depicts the energy spectrum with the bulk bands
and the edge states. For 1/6 filling only the lowest bulk band is
filled, which is well separated from the next band. We have
calculated the Chern number ($C$), which characterizes the quantum
Hall effect by giving the number of counter propagating edge state
pairs inside a bulk energy gap. We have found that $C=6$ for the
lowest gap, which is in accordance with the observation that there
is one edge state pair per spin component. It follows that an
elementary flux of $\Phi_0=\pi/3$ is attached to every spinon (the
fractionalized quasi-particle excitation, created by the operators
$c_{i,\alpha}^\dagger$), which in effect renders their statistics
to an anyonic one. On the other hand it provides explanation for
the $\Phi=2\pi/3$ flux per plaquette of the ground state: since
each site belongs to 3 plaquettes, every plaquette contains 2
spinons leading to $\Phi=2\Phi_0$, as expected.

The low-energy fluctuations of the mean-field theory are given by
the phase fluctuations of the mean-field amplitude,
$\chi_{ij}=\chi_{ij}^{\text{mf}}\,e^{ia_{ij}}$, and by the lowest
energy spinon excitations. The mean-field phase fluctuation
$a_{ij}$ is a gauge field with the transformation property
$a_{ij}\rightarrow a_{ij}+\theta_i-\theta_j$. The low energy
spinon excitations are coming from the vicinity of the energy
maxima of the valence band, and from the energy minima of the
conduction band. Since there are 6 such points, the low-energy
dynamics is governed by spinons with 6 flavors interacting with a
U(1) gauge field. By integrating out the high energy (gapped)
spinon fields the resulting low energy theory is described by the
following Lagrangian density in the continuum limit:
\begin{multline}
\label{eq:lagr}
\mathcal{L}=\frac{1}{8\pi q^2}(\mathbf{e}^2-vb^2) - \frac{C}{4\pi}\epsilon^{\mu\nu\lambda}a_\mu\partial_\nu a_\lambda\\
+\sum_{l=1}^{6}\left[-i c_{l,\alpha}^\dagger(\partial_t-ia_0)c_{l,\alpha}+\frac{1}{2 m_s}c_{l,\alpha}^\dagger(\partial_i+i a_i)^2c_{l,\alpha}\right],
\end{multline}
where the constant $q^2$ arises from the integration of the spinon
fields and is in the order of the spinon gap $g$. The fields
$\mathbf{e}$, and $b$, are the artificial electric and magnetic
fields, respectively, obtained from the scalar and vector
potentials $a_0$ and $a_i$ in the usual way. The constant $C=6$ is
the Chern number, and the speed of sound $v$ is proportional to
$1/g\sim U/t^2$. In the spinon part of the Lagrangian $m_s$ stands
for the effective mass of the spinons, which can be obtained from
the curvature of the spinon dispersion around the 6 minimas. Due
to the Chern-Simons term, the gauge bosons mediate only a short
range interaction between the spinons, and the mean-field solution
is stable \cite{wenbook}. Therefore, the low energy properties of
the system can be characterized by U(1) gauge field coupled to the
spinons as matter fields. The corresponding gauge theory described
by the Lagrangian \eqref{eq:lagr} can be thus simulated with
ultracold spin-5/2 fermions loaded into hexagonal lattice.

Ultracold alkaline earth atoms are
produced routinely nowadays. A honeycomb optical lattice can also
be created by sophisticated laser configurations
\cite{soltan-panahi11a}.  Another, even cleaner experimental
implementation would be to use the holographic methods of Greiner
{\it et al.}, or Esslinger {\it et al.}, where an arbitrary two
dimensional lattice potential can be created with the help of an
optical imaging system \cite{bakr09a,zimmermann11a}. Detecting the
CSL phase, or the emerging dynamical gauge theory
is not straightforward, but possible. For example, one can measure
nearest neighbor pair correlations \cite{greif11a}, but there is
only access to $|\chi_{ij}|$. In fact, according to the Elitzur's theorem
non gauge invariant quantities, such as $\chi_{ij}$,
average to zero \cite{baskaran88a,elitzur75a}. A gauge invariant
quantity sensitive to chirality and possible to measure is the phase of a loop, which can be
detected directly by measuring 3-spin correlations:
$\vec{S}_i\cdot(\vec{S}_j\times\vec{S}_k)$. It's nonzero value
witnesses for the chiral nature of the spin liquid phase
\cite{wen89a}. Finally, and more importantly, one can measure
in experiments, using for instance spin polarization spectroscopy \cite{hammerer10a}, the spin structure factor $S^{zz}(i,j;t)=\langle
S^z_i(t)S^z_j(0)\rangle$ at site $i$ and $j$ at time $t$ and zero,
respectively. This quantity  can be expressed with the help of the
four point spinon Green's functions, and in the RPA approximation
is given by
\begin{equation}
\label{eq:strfacRPA}
S^{zz}(\mathbf{k},\omega)=F^{z}_{\alpha\beta}F^z_{\beta\alpha}\frac{\Pi(\mathbf{k},\omega)}{1-6\,D(\mathbf{k},\omega)\,\Pi(\mathbf{k},\omega)},
\end{equation}
with $F^z_{\alpha\beta}$ being the $z$-component of the spin matrix in the 5/2 representation. The quantity
$\Pi(\mathbf{k},\omega)$ is the spinon polarization function,
which to lowest order is simply the contribution of the bubble
diagram, and $D(\mathbf{k},\omega)$ is the photon propagator up to
a numeric matrix for contracting the space time indices. Due to
the hybridization of the gauge field and spinon
propagators one can find resonances in the magnetic response
function \eqref{eq:strfacRPA} belonging to the spinon and to the
gauge field excitations. Therefore magnetic response measurements
are suitable to reveal the gauge structure of low-lying
excitations, and the chiral nature of the ground state.

We have studied the one particle per site Mott
insulator phases of spin-5/2 ultracold alkaline earth atoms in the
honeycomb lattice. We have found that the ground state is a chiral
spin liquid state with broken time reversal symmetry. Thanks to
the finite gap appearing in the spinon spectrum, we have
integrated out the high energy spinon fields and arrived to a
dynamical U(1) gauge field theory with a Chern-Simons term. This
gauge theory describes the spin fluctuations of the system, and
therefore the gauge field dynamics can be accessed experimentally
with the help of spin response measurements.

It is a pleasure for us to acknowledge the useful discussions with
J\'anos Asb\'oth. We acknowledge funding from the Spanish MEC
projects TOQATA (FIS2008-00784), QOIT (Consolider Ingenio 2010),
ERC Advanced Grant QUAGATUA, EU STREP NAMEQUAM, NSF of Hungary
(OTKA) (T077629 and 68340) and Alexander von Humboldt Foundation
(M. L.).


\begin{thebibliography}{29}
\expandafter\ifx\csname natexlab\endcsname\relax\def\natexlab#1{#1}\fi
\expandafter\ifx\csname bibnamefont\endcsname\relax
  \def\bibnamefont#1{#1}\fi
\expandafter\ifx\csname bibfnamefont\endcsname\relax
  \def\bibfnamefont#1{#1}\fi
\expandafter\ifx\csname citenamefont\endcsname\relax
  \def\citenamefont#1{#1}\fi
\expandafter\ifx\csname url\endcsname\relax
  \def\url#1{\texttt{#1}}\fi
\expandafter\ifx\csname urlprefix\endcsname\relax\def\urlprefix{URL }\fi
\providecommand{\bibinfo}[2]{#2}
\providecommand{\eprint}[2][]{\url{#2}}

\bibitem[{\citenamefont{Lewenstein et~al.}(2007)}]{lewenstein07a}
\bibinfo{author}{\bibfnamefont{M.}~\bibnamefont{Lewenstein}}
  \bibnamefont{et~al.}, \bibinfo{journal}{Advances in Physics}
  \textbf{\bibinfo{volume}{56}}, \bibinfo{pages}{243} (\bibinfo{year}{2007}).

\bibitem[{\citenamefont{Fukuhara et~al.}(2007)\citenamefont{Fukuhara, Takasu,
  Kumakura, and Takahashi}}]{fukuhara07a}
\bibinfo{author}{\bibfnamefont{T.}~\bibnamefont{Fukuhara}},
  \bibinfo{author}{\bibfnamefont{Y.}~\bibnamefont{Takasu}},
  \bibinfo{author}{\bibfnamefont{M.}~\bibnamefont{Kumakura}}, \bibnamefont{and}
  \bibinfo{author}{\bibfnamefont{Y.}~\bibnamefont{Takahashi}},
  \bibinfo{journal}{Phys. Rev. Lett.} \textbf{\bibinfo{volume}{98}},
  \bibinfo{pages}{030401} (\bibinfo{year}{2007}).

\bibitem[{\citenamefont{DeSalvo et~al.}(2010)}]{desalvo10a}
\bibinfo{author}{\bibfnamefont{B.~J.} \bibnamefont{DeSalvo}}
  \bibnamefont{et~al.}, \bibinfo{journal}{Phys. Rev. Lett.}
  \textbf{\bibinfo{volume}{105}}, \bibinfo{pages}{030402}
  (\bibinfo{year}{2010}).

\bibitem[{\citenamefont{Taie et~al.}(2010)}]{taie10a}
\bibinfo{author}{\bibfnamefont{S.}~\bibnamefont{Taie}} \bibnamefont{et~al.},
  \bibinfo{journal}{Phys. Rev. Lett.} \textbf{\bibinfo{volume}{105}},
  \bibinfo{pages}{190401} (\bibinfo{year}{2010}).

\bibitem[{\citenamefont{Gorshkov et~al.}(2010)}]{gorshkov10a}
\bibinfo{author}{\bibfnamefont{A.~V.} \bibnamefont{Gorshkov}}
  \bibnamefont{et~al.}, \bibinfo{journal}{Nature Physics}
  \textbf{\bibinfo{volume}{6}}, \bibinfo{pages}{289} (\bibinfo{year}{2010}).

\bibitem[{\citenamefont{Lee et~al.}(2006)\citenamefont{Lee, Nagaosa, and
  Wen}}]{lee06a}
\bibinfo{author}{\bibfnamefont{P.~A.} \bibnamefont{Lee}},
  \bibinfo{author}{\bibfnamefont{N.}~\bibnamefont{Nagaosa}}, \bibnamefont{and}
  \bibinfo{author}{\bibfnamefont{X.-G.} \bibnamefont{Wen}},
  \bibinfo{journal}{Rev. Mod. Phys.} \textbf{\bibinfo{volume}{78}},
  \bibinfo{pages}{17} (\bibinfo{year}{2006}).

\bibitem[{\citenamefont{Cirac et~al.}(2010)\citenamefont{Cirac, Maraner, and
  Pachos}}]{cirac10a}
\bibinfo{author}{\bibfnamefont{J.~I.} \bibnamefont{Cirac}},
  \bibinfo{author}{\bibfnamefont{P.}~\bibnamefont{Maraner}}, \bibnamefont{and}
  \bibinfo{author}{\bibfnamefont{J.~K.} \bibnamefont{Pachos}},
  \bibinfo{journal}{Phys. Rev. Lett.} \textbf{\bibinfo{volume}{105}},
  \bibinfo{pages}{190403} (\bibinfo{year}{2010}).

\bibitem[{\citenamefont{Kapit and Mueller}(2011)}]{kapit11a}
\bibinfo{author}{\bibfnamefont{E.}~\bibnamefont{Kapit}} \bibnamefont{and}
  \bibinfo{author}{\bibfnamefont{E.}~\bibnamefont{Mueller}},
  \bibinfo{journal}{Phys. Rev. A} \textbf{\bibinfo{volume}{83}},
  \bibinfo{pages}{033625} (\bibinfo{year}{2011}).

\bibitem[{\citenamefont{Wen}(2004)}]{wenbook}
\bibinfo{author}{\bibfnamefont{X.-G.} \bibnamefont{Wen}},
  \emph{\bibinfo{title}{Quantum Field Theory of Many-Body systems}}
  (\bibinfo{publisher}{Oxford University Press}, \bibinfo{year}{2004}).

\bibitem[{\citenamefont{Hermele et~al.}(2009)\citenamefont{Hermele, Gurarie,
  and Rey}}]{hermele09a}
\bibinfo{author}{\bibfnamefont{M.}~\bibnamefont{Hermele}},
  \bibinfo{author}{\bibfnamefont{V.}~\bibnamefont{Gurarie}}, \bibnamefont{and}
  \bibinfo{author}{\bibfnamefont{A.~M.} \bibnamefont{Rey}},
  \bibinfo{journal}{Phys. Rev. Lett.} \textbf{\bibinfo{volume}{103}},
  \bibinfo{pages}{135301} (\bibinfo{year}{2009}).

\bibitem[{\citenamefont{Wen et~al.}(1989)\citenamefont{Wen, Wilczek, and
  Zee}}]{wen89a}
\bibinfo{author}{\bibfnamefont{X.~G.} \bibnamefont{Wen}},
  \bibinfo{author}{\bibfnamefont{F.}~\bibnamefont{Wilczek}}, \bibnamefont{and}
  \bibinfo{author}{\bibfnamefont{A.}~\bibnamefont{Zee}},
  \bibinfo{journal}{Phys. Rev. B} \textbf{\bibinfo{volume}{39}},
  \bibinfo{pages}{11413} (\bibinfo{year}{1989}).

\bibitem[{\citenamefont{Meng et~al.}(2010)\citenamefont{Meng, Lang, Wessel,
  Assaad, and Muramatsu}}]{meng10a}
\bibinfo{author}{\bibfnamefont{Z.~Y.} \bibnamefont{Meng}},
  \bibinfo{author}{\bibfnamefont{T.~C.} \bibnamefont{Lang}},
  \bibinfo{author}{\bibfnamefont{S.}~\bibnamefont{Wessel}},
  \bibinfo{author}{\bibfnamefont{F.~F.} \bibnamefont{Assaad}},
  \bibnamefont{and}
  \bibinfo{author}{\bibfnamefont{A.}~\bibnamefont{Muramatsu}},
  \bibinfo{journal}{Nature} \textbf{\bibinfo{volume}{464}},
  \bibinfo{pages}{847} (\bibinfo{year}{2010}).

\bibitem[{\citenamefont{Black-Schaffer and Doniach}(2007)}]{black-schaffer07a}
\bibinfo{author}{\bibfnamefont{A.~M.} \bibnamefont{Black-Schaffer}}
  \bibnamefont{and} \bibinfo{author}{\bibfnamefont{S.}~\bibnamefont{Doniach}},
  \bibinfo{journal}{Phys. Rev. B} \textbf{\bibinfo{volume}{75}},
  \bibinfo{pages}{134512} (\bibinfo{year}{2007}).

\bibitem[{\citenamefont{Pathak et~al.}(2010)\citenamefont{Pathak, Shenoy, and
  Baskaran}}]{pathak10a}
\bibinfo{author}{\bibfnamefont{S.}~\bibnamefont{Pathak}},
  \bibinfo{author}{\bibfnamefont{V.~B.} \bibnamefont{Shenoy}},
  \bibnamefont{and} \bibinfo{author}{\bibfnamefont{G.}~\bibnamefont{Baskaran}},
  \bibinfo{journal}{Phys. Rev. B} \textbf{\bibinfo{volume}{81}},
  \bibinfo{pages}{085431} (\bibinfo{year}{2010}).

\bibitem[{\citenamefont{Li}(2011)}]{li11a}
\bibinfo{author}{\bibfnamefont{T.}~\bibnamefont{Li}} (\bibinfo{year}{2011}),
  \eprint{arXiv:1101.1352}.

\bibitem[{\citenamefont{Machida et~al.}(2009)\citenamefont{Machida, Nakatsuji,
  Onoda, Tayama, and Sakakibara}}]{machida09a}
\bibinfo{author}{\bibfnamefont{Y.}~\bibnamefont{Machida}},
  \bibinfo{author}{\bibfnamefont{S.}~\bibnamefont{Nakatsuji}},
  \bibinfo{author}{\bibfnamefont{S.}~\bibnamefont{Onoda}},
  \bibinfo{author}{\bibfnamefont{T.}~\bibnamefont{Tayama}}, \bibnamefont{and}
  \bibinfo{author}{\bibfnamefont{T.}~\bibnamefont{Sakakibara}},
  \bibinfo{journal}{Nature} \textbf{\bibinfo{volume}{463}},
  \bibinfo{pages}{210} (\bibinfo{year}{2009}).

\bibitem[{\citenamefont{Koch et~al.}(2010)\citenamefont{Koch, Houck, Hur, and
  Girvin}}]{koch10a}
\bibinfo{author}{\bibfnamefont{J.}~\bibnamefont{Koch}},
  \bibinfo{author}{\bibfnamefont{A.~A.} \bibnamefont{Houck}},
  \bibinfo{author}{\bibfnamefont{K.~L.} \bibnamefont{Hur}}, \bibnamefont{and}
  \bibinfo{author}{\bibfnamefont{S.~M.} \bibnamefont{Girvin}},
  \bibinfo{journal}{Phys. Rev. A} \textbf{\bibinfo{volume}{82}},
  \bibinfo{pages}{043811} (\bibinfo{year}{2010}).

\bibitem[{\citenamefont{Wen}(2002)}]{wen02a}
\bibinfo{author}{\bibfnamefont{X.-G.} \bibnamefont{Wen}},
  \bibinfo{journal}{Phys. Rev. B} \textbf{\bibinfo{volume}{65}},
  \bibinfo{pages}{165113} (\bibinfo{year}{2002}).

\bibitem[{\citenamefont{Anderson}(1973)}]{anderson73a}
\bibinfo{author}{\bibfnamefont{P.}~\bibnamefont{Anderson}},
  \bibinfo{journal}{Materials Research Bulletin} \textbf{\bibinfo{volume}{8}},
  \bibinfo{pages}{153 } (\bibinfo{year}{1973}).

\bibitem[{\citenamefont{Fazekas and Anderson}(1974)}]{fazekas74a}
\bibinfo{author}{\bibfnamefont{P.}~\bibnamefont{Fazekas}} \bibnamefont{and}
  \bibinfo{author}{\bibfnamefont{P.~W.} \bibnamefont{Anderson}},
  \bibinfo{journal}{Philos. Mag.} \textbf{\bibinfo{volume}{30}},
  \bibinfo{pages}{432} (\bibinfo{year}{1974}).

\bibitem[{\citenamefont{Marston and Affleck}(1989)}]{marston89a}
\bibinfo{author}{\bibfnamefont{J.~B.} \bibnamefont{Marston}} \bibnamefont{and}
  \bibinfo{author}{\bibfnamefont{I.}~\bibnamefont{Affleck}},
  \bibinfo{journal}{Phys. Rev. B} \textbf{\bibinfo{volume}{39}},
  \bibinfo{pages}{11538} (\bibinfo{year}{1989}).

\bibitem[{\citenamefont{Szirmai and Lewenstein}(2011)}]{szirmai11a}
\bibinfo{author}{\bibfnamefont{E.}~\bibnamefont{Szirmai}} \bibnamefont{and}
  \bibinfo{author}{\bibfnamefont{M.}~\bibnamefont{Lewenstein}},
  \bibinfo{journal}{EPL} \textbf{\bibinfo{volume}{93}}, \bibinfo{pages}{66005}
  (\bibinfo{year}{2011}).

\bibitem[{\citenamefont{Soltan-Panahi et~al.}(2011)}]{soltan-panahi11a}
\bibinfo{author}{\bibfnamefont{P.}~\bibnamefont{Soltan-Panahi}}
  \bibnamefont{et~al.}, \bibinfo{journal}{Nat. Phys.}
  \textbf{\bibinfo{volume}{7}}, \bibinfo{pages}{434} (\bibinfo{year}{2011}).

\bibitem[{\citenamefont{Bakr et~al.}(2009)}]{bakr09a}
\bibinfo{author}{\bibfnamefont{W.~S.} \bibnamefont{Bakr}} \bibnamefont{et~al.},
  \bibinfo{journal}{Nature} \textbf{\bibinfo{volume}{462}}, \bibinfo{pages}{74}
  (\bibinfo{year}{2009}).

\bibitem[{\citenamefont{Zimmermann et~al.}(2011)}]{zimmermann11a}
\bibinfo{author}{\bibfnamefont{B.}~\bibnamefont{Zimmermann}}
  \bibnamefont{et~al.}, \bibinfo{journal}{New J. Phys.}
  \textbf{\bibinfo{volume}{13}}, \bibinfo{pages}{043007}
  (\bibinfo{year}{2011}).

\bibitem[{\citenamefont{Greif et~al.}(2011)}]{greif11a}
\bibinfo{author}{\bibfnamefont{D.}~\bibnamefont{Greif}} \bibnamefont{et~al.},
  \bibinfo{journal}{Phys. Rev. Lett.} \textbf{\bibinfo{volume}{106}},
  \bibinfo{pages}{145302} (\bibinfo{year}{2011}).

\bibitem[{\citenamefont{Baskaran and Anderson}(1988)}]{baskaran88a}
\bibinfo{author}{\bibfnamefont{G.}~\bibnamefont{Baskaran}} \bibnamefont{and}
  \bibinfo{author}{\bibfnamefont{P.~W.} \bibnamefont{Anderson}},
  \bibinfo{journal}{Phys. Rev. B} \textbf{\bibinfo{volume}{37}},
  \bibinfo{pages}{580} (\bibinfo{year}{1988}).

\bibitem[{\citenamefont{Elitzur}(1975)}]{elitzur75a}
\bibinfo{author}{\bibfnamefont{S.}~\bibnamefont{Elitzur}},
  \bibinfo{journal}{Phys. Rev. D} \textbf{\bibinfo{volume}{12}},
  \bibinfo{pages}{3978} (\bibinfo{year}{1975}).

\bibitem[{\citenamefont{Hammerer et~al.}(2010)\citenamefont{Hammerer,
  S\o{}rensen, and Polzik}}]{hammerer10a}
\bibinfo{author}{\bibfnamefont{K.}~\bibnamefont{Hammerer}},
  \bibinfo{author}{\bibfnamefont{A.~S.} \bibnamefont{S\o{}rensen}},
  \bibnamefont{and} \bibinfo{author}{\bibfnamefont{E.~S.}
  \bibnamefont{Polzik}}, \bibinfo{journal}{Rev. Mod. Phys.}
  \textbf{\bibinfo{volume}{82}}, \bibinfo{pages}{1041} (\bibinfo{year}{2010}).

\end{thebibliography}
\end{document}